\documentclass{PoS}

\graphicspath{{./images/}}
\usepackage{graphbox}

\title{LATTES: a novel detector concept for a gamma-ray experiment in the Southern hemisphere}

\ShortTitle{LATTES}

\author{Pedro Assis, \speaker{Ruben Concei\c{c}\~{a}o}\thanks{Acknowledges the financial support of Funda\c{c}\~ao para a Ci\^encia e Tecnologia, FCT-Portugal.} , M\'ario Pimenta, Bernardo Tom\'e\\
        Laborat\'orio de Instrumenta\c{c}\~{a}o e F\'isica Experimental de Partículas - LIP and Instituto Superior T\'ecnico - IST, Universidade de Lisboa - UL, Portugal\\
        E-mail: \email{ruben@lip.pt}}

\author{Alberto Blanco, Luis Lopes\\
        Laborat\'orio de Instrumenta\c{c}\~{a}o e F\'isica Experimental de Partículas - LIP, Coimbra, Portugal\\}
        
\author{Paulo Fonte\\
        Coimbra Polytechnic - ISEC and 
        Laborat\'orio de Instrumenta\c{c}\~{a}o e F\'isica Experimental de Partículas - LIP, Coimbra, Portugal\\}
        
 \author{Ulisses Barres de Almeida, Ronald Shellard\\
Centro Brasileiro de Pesquisas F\'isicas - CBPF, Rio de Janeiro, Brazil\\}
   
    \author{Benedetto D'Ettorre Piazzoli\\
Universit\`a di Napoli Federico II and INFN Roma Tor Vergata, Italy\\}

    \author{Alessandro De Angelis\\
INFN Padova, Universit\`a di Udine, Italy and Laborat\'orio de Instrumenta\c{c}\~{a}o e F\'isica Experimental de Partículas - LIP, Instituto Superior T\'ecnico - IST, Universidade de Lisboa - UL, Portugal\\}

    \author{Michele Doro\\
Department of Physics and Astronomy,
University of Padova and INFN Padova, E-35131 Padova, Italy\\}

    \author{Giorgio Matthiae\\
INFN and Universit\`a di Roma Tor Vergata, Roma, Italy\\}

\abstract{The Large Array Telescope for Tracking Energetic Sources (LATTES), is a novel concept for an array of hybrid EAS array detectors, composed of a Resistive Plate Counter array coupled to a Water Cherenkov Detector, planned to cover gamma rays from less than 100 GeV up to 100 TeVs. This experiment, to be installed at high altitude in South America, could cover the existing gap in sensitivity between satellite and ground arrays.

The low energy threshold, large duty cycle and wide field of view of LATTES makes it a powerful tool to detect transient phenomena and perform long term observations of variable sources. Moreover, given its characteristics, it would be fully complementary to the planned Cherenkov Telescope Array (CTA) as it would be able to issue alerts.

In this talk, a description of its main features and capabilities, as well as results on its expected performance, and sensitivity, will be presented.}

\FullConference{35th International Cosmic Ray Conference --- ICRC2017\\
		10--20 July, 2017\\
		Bexco, Busan, Korea}

\begin{document}

\section{Introduction}

Gamma rays are one of the best probes to study our surrounding Universe. Due to their neutral nature, gamma rays are not deflected by the interstellar magnetic field which means that they can be used to pin-point the emitting astrophysical sources. Very high-energy gamma rays, comprised in an energy range between $\sim 100\,$GeV and $\sim 100\,$TeV are particularly interesting as they allow us to investigate some of the most extreme phenomena.

At energies up to $\sim 100\,$GeV, this radiation can be detected directly using instruments placed in artificial satellites, as for instance Fermi. Above this energies the gamma ray flux becomes too small and indirect methods prevail. These methods take advantage of the interaction of the gamma-ray with Earth's atmosphere which produces a cascade of particles designated usually as Extensive Air Showers (EAS).

At ground there are two main techniques: Imaging Atmospheric Cherenkov Telescope Arrays (IACTAs) and EAS arrays. The former technique measures the Cherenkov light produce by relativistic shower particles while the later measures the secondaries particles of the EAS that reach the ground.  Both techniques are placed at high altitude to minimize the atmosphere's attenuation. There are advantages and disadvantages to both approaches. IACTs have a better energy and geometry reconstruction resolution but can only survey small portions of the sky. Moreover, they have a limited duty cycle. This makes them perfect to study astrophysical sources but not to look for transient phenomena.

Presently there is no wide Field Of View (FoV) experiment operating in the Southern hemisphere, nor able to cover the gap between satellite and ground array gamma-ray experiments.

A low energy threshold and a large duty cycle wide FoV experiment would be fully complementary to planned project such as the Cherenkov Telescope Array (CTA), as it  could issue alert of transient phenomena. Moreover, such an experiment is able to perform long term observations of variable sources and search for emissions from extended regions, arising from phenomena such as the Fermi bubbles or dark matter annihilations from the centre of our galaxy. 
 
Hence, we propose a novel hybrid detector to be installed at $\sim 5200\,$m a.s.l. which has an improved sensitivity at the 100 GeV energy region.
This manuscript is organised as follows: in section 2 we describe the detector and the layout of the experiment. In section 3 we discussed the its performance and the achieved preliminary sensitivity. Finally, we end with a summary.

\section{Detector description}

\begin{figure}
     \centering
     \includegraphics[align=c, width=.4\textwidth]{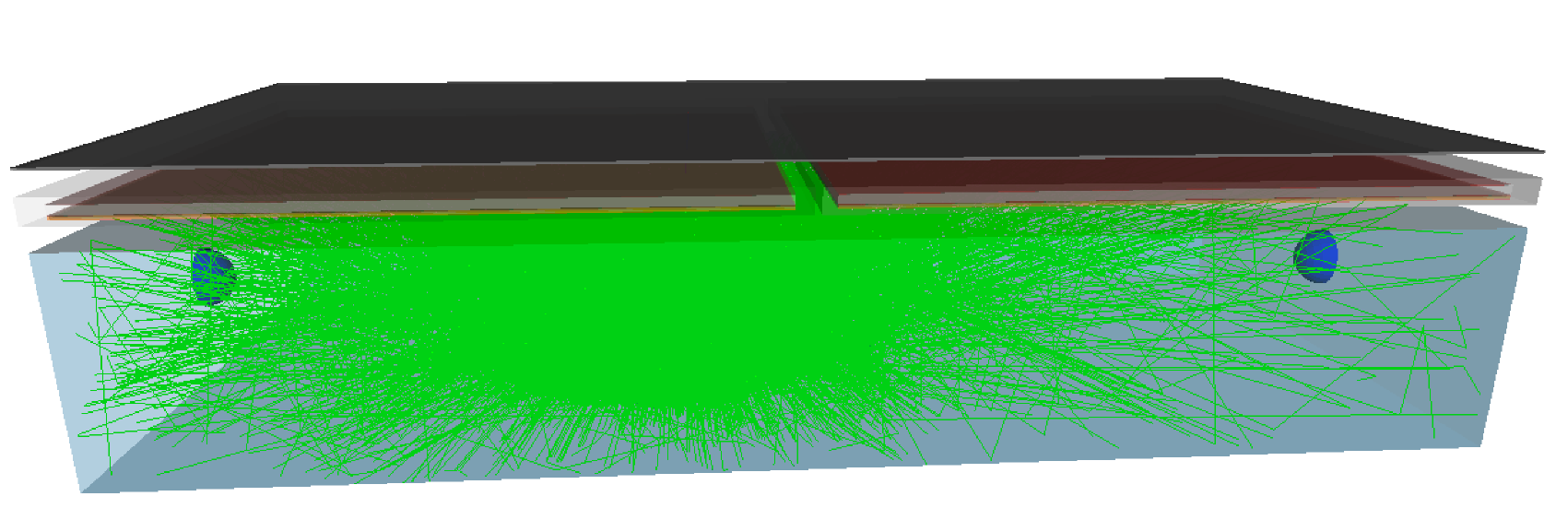}
     \hspace*{5mm}
     \includegraphics[align=c, width=.4\textwidth]{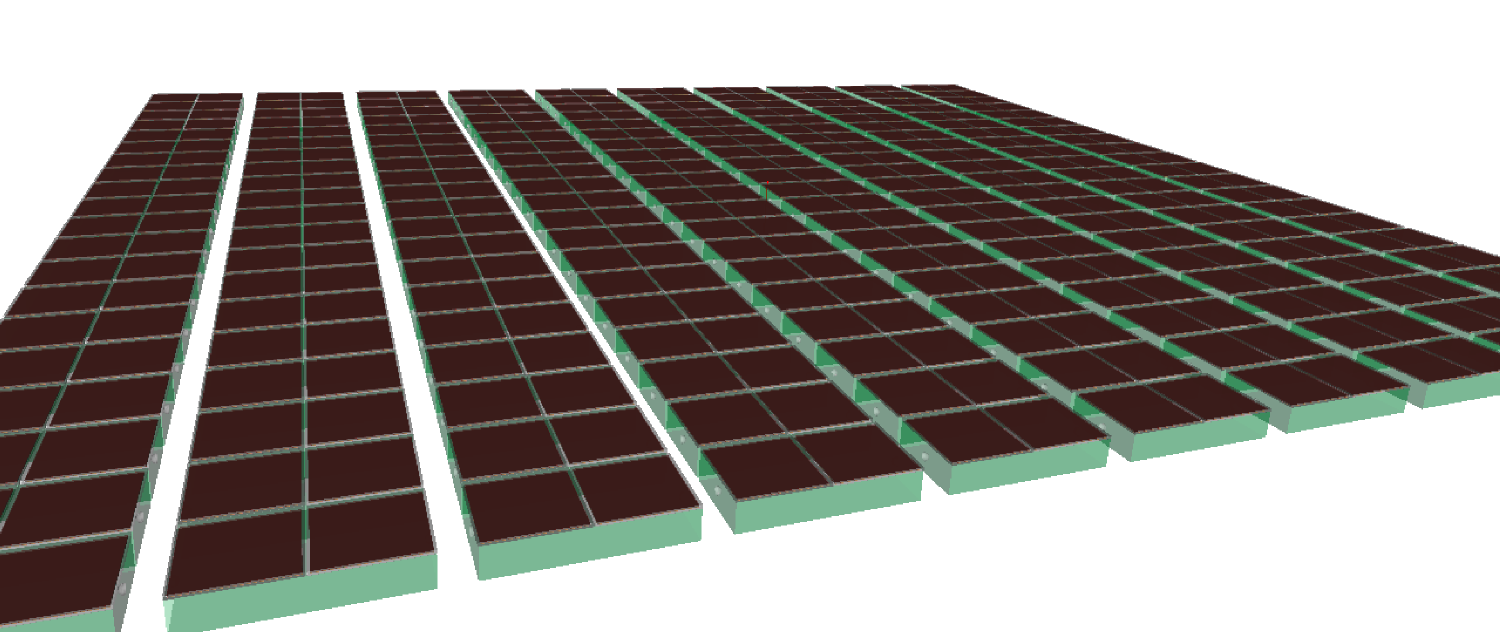}
     \caption{(left) Basic detector station, with one WCD covered with RPCs and a thin slab of lead. The green lines show the tracks of the Cherenkov photons produced by the electron and positron from the conversion of a photon in the lead slab. (right) Layout of the compact core array. }
  \label{fig:LATTES}
\end{figure}

Lowering the energy threshold requires that one would be able to trigger on the shower secondary photons which are more numerous by a factor of 5-7 than secondary charged particles. As such, we propose to build a dense array with an area of $20\,000\,{\rm m^2}$ constituted by modular hybrid detectors (see figure~\ref{fig:LATTES} (left)). Each station is composed by two low-cost Resistive Plate Chambers (RPC) on top of a Water Cherenkov Detector (WCD), as shown in figure~\ref{fig:LATTES} (right). Each RPC has 16 charge collecting pads covering a total area of $1.5 \times 1.5\,{\rm m^2}$ . The WCD has a rectangular structure with dimensions $3\times1.5\times 0.5\,{\rm m^3}$. The signals are read by two photomultipliers (PMTs) at both ends of the smallest vertical face of the WCD. This detector concept ensures the ability to trigger at low energies while maintaining a reasonable energy and geometry reconstruction accuracy, which is essential to distinguish the gamma-ray showers from the ones initiated by cosmic rays.

The geometry reconstruction can be further improved by adding on the top of the RPCs a thin lead plate (5.6 mm). This allows to convert secondary photons, which have a stronger correlation with the shower axis while removing low energy electrons who have poorer correlation due to multiple scattering in the atmosphere.

Hence, the RPCs contributes, with its high segmentation and time resolution, which is essential for the shower geometric reconstruction, while the WCD provides a calorimetric measurement of the shower secondary particles lowering effectively the experiment energy threshold. 


\section{Detector performance}

The performance of this detector has been assessed using an end-to-end realistic Monte Carlo simulation. Extensive Air Showers (EAS) have been simulated using CORSIKA (COsmic Ray Simulations for KAscade)~\cite{corsika}, while the detector response has been treated with Geant4~\cite{geant4}. A total of $5\times10^6$ gamma and proton showers have been simulated with energies between $10\,$GeV and $300\,$TeV. The simulations were generated uniformly in logarithm of the energy and afterwards weighted accordingly to the corresponding particle fluxes. The zenith angle for gammas was fixed to $10^\circ$, while for protons the range was between 5 and 15 degrees. An altitude of 5200 a.s.l. was chosen to investigate the experiment performance.

\subsection{Trigger and effective area}

\begin{figure}
     \centering
     \includegraphics[align=c, width=.7\textwidth]{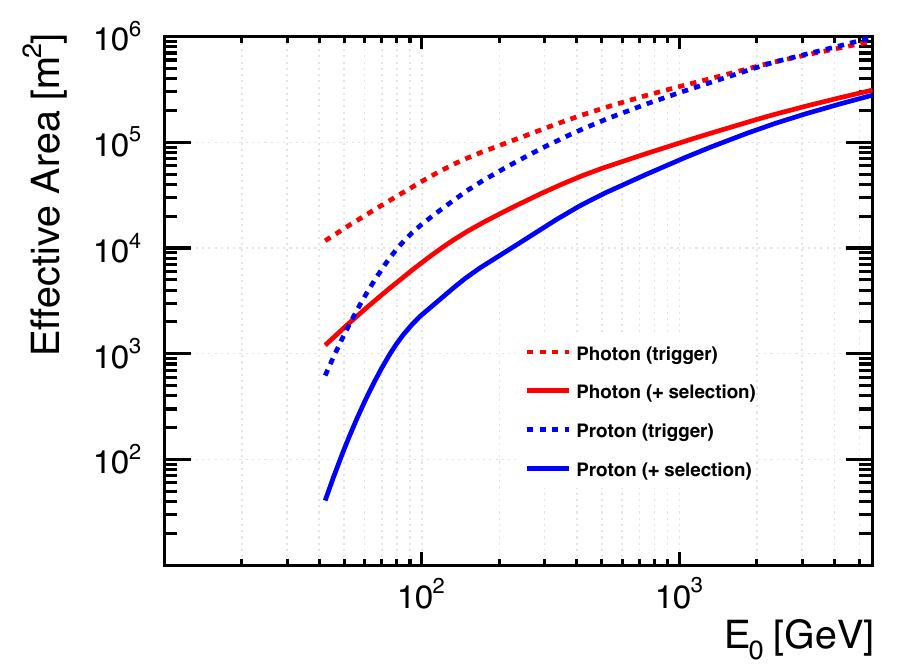}
     \caption{Effective area at trigger level (dashed curves) and after the quality cuts used for the shower geometry reconstruction and gamma/hadron discrimination (solid curves), separately for gamma-ray and proton-initiated showers. The effective area is shown as a function of the simulated energy, $E_0$.}
  \label{fig:EffArea}
\end{figure}

As stated before, the trigger is ensured by the WCD. As such a station is considered to be triggered if each PMT collects at least 5 photoelectrons. To have an event one should have at least 3 active WCD stations.
The effective area has been computed using simulations and is shown in figure~\ref{fig:EffArea}. From this figure it is possible to see that, even after quality cuts, one still has for gamma primaries with an energy of 100 GeV an effective area of about $10^4\,{\rm m^2}$. The applied quality cuts shall be described in the next sections.

\subsection{Energy reconstruction}

\begin{figure}
     \centering
     \includegraphics[align=c, width=.4\textwidth]{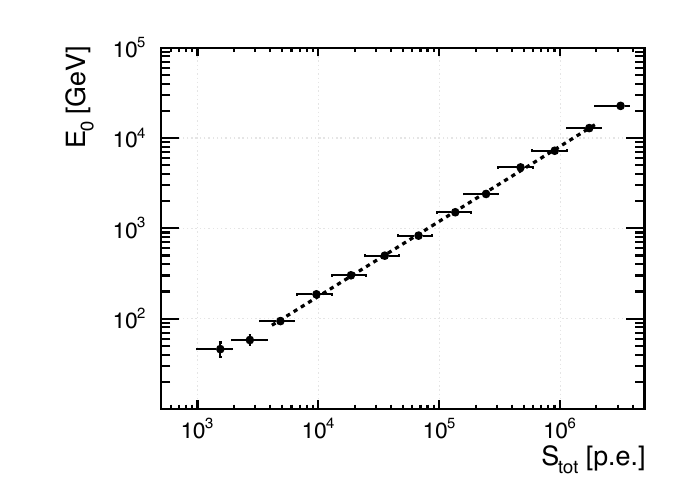}
     \hspace*{5mm}
     \includegraphics[align=c, width=.4\textwidth]{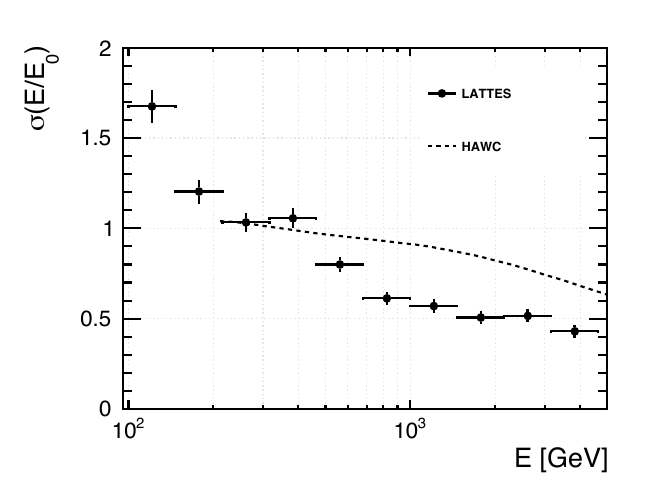}
     \caption{(left) Calibration curve between the total signal recorded by all WCD stations and the simulated energy. This calibration curve was built assuming photons with a spectral energy distribution as for Crab nebula(right) Resolution in the reconstructed energy for the same sample. The HAWC curve is shown for comparison~\cite{HAWC2017}. }
  \label{fig:Erec}
\end{figure}

The energy estimation has been obtained using the signal measured by the WCDs. A calibration was built relating the total signal, $S_{tot}$. recorded in all WCD stations for each event with the true energy of the primary gamma, $E_0$. From this curve it is possible to evaluate the reconstructed energy, $E_{rec}$, for each shower and assess the energy resolution for this detector. In figure~\ref{fig:Erec} (right) it can be seen that the energy resolution improves, as expected, with the increase of the shower energy, while at lower energy it degrades considerably. The latter is due to shower-to-shower fluctuations. A comparison with HAWC results, presented in~\cite{HAWC2017}, is also shown in this figure. It can be seen that the LATTES energy resolution is better above $500\,$GeV and is comparable below. The reason why LATTES performs so well at high energies is related to its higher segmentation and compactness.

\subsection{Geometry reconstruction}

\begin{figure}
     \centering
     \includegraphics[align=c, width=.4\textwidth]{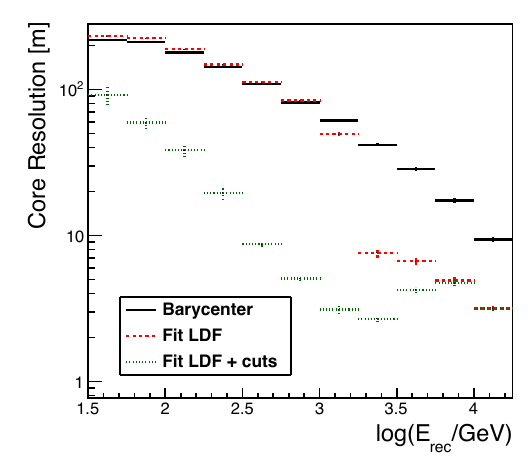}
     \hspace*{5mm}
     \includegraphics[align=c, width=.5\textwidth]{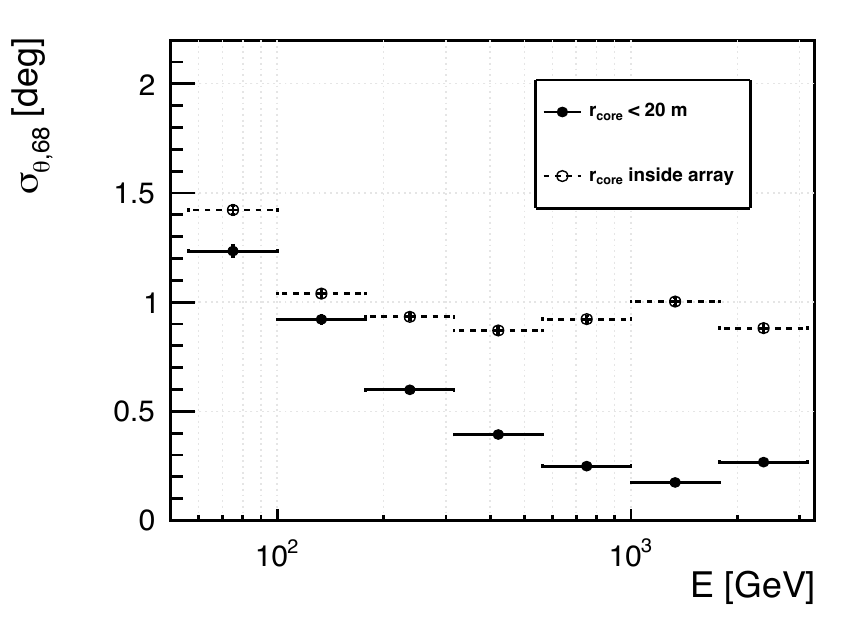}
     \caption{(left) Resolution of the reconstructed shower core for different procedures (see legend for details). (right) Angular resolution for gamma-ray primaries with zenith angle $\theta = 10^\circ$, as a function of the reconstructed energy. }
  \label{fig:geom}
\end{figure}

The reconstruction of the shower core was done fitting a gamma average 2D LDF to the WCD station signal with a procedure similar to the one discussed in~\cite{HAWC2017}. To require a clear maximum inside of the array, a cut on the topology of the event has been done. In figure~\ref{fig:geom} (left) it can be seen that the shower core can be reconstructed with an accuracy better than $10\,$m for gamma induced showers with energies above 300 GeV.

The shower geometry reconstruction is being done taking advantage of the RPC segmentation and fast timing. It was considered a time resolution of $1\,$ns. The primary direction is obtained using a shower front plane model that has as ingredients the position and time of the recorded hits in the RPC.
In order to improve the quality of the reconstruction it is required that the event has at least 10 hits on the RPCs pads. Only RPCs on top of active WCDs are considered to apply this cut. Moreover, late arrival particles are also discarded. The reconstructed angle was compared to the simulated one, and we calculate the 68\% containment angle, $\sigma_{\theta,68}$. In figure~\ref{fig:geom} (right) it is shown $\sigma_{\theta,68}$ as a function of the shower reconstructed energy for two conditions: when the reconstructed core is contained in the array and when the reconstructed core is required to be at a distance smaller than $20\,$m from the array center. From this figure it can be seen that  at energies around 100 GeV, a reasonable resolution, better that $1.5^\circ$, can be achieved.

Moreover, it is noticeable from this plot that  at higher energies the events reconstructed near to the array center have a better resolution. This clearly shows that the front plane model is not good enough to deal with the border effects of the array. This simplistic approach should be substituted by a shower conic fit, which accounts for the shower front curvature.

\subsection{Gamma-hadron discrimination}

The shower characteristics may be used to distinguish a pure electromagnetic shower, induced by a gamma-ray, from a shower generated by a hadronic primary. Here, we attempt to identify observables able to distinguish the two kind of primaries based solely on the WCD signal. Two promising variables, inspired in the procedures followed in~\cite{HAWC2017} were found:
\begin{itemize}
\item Sum of the signal of all WCD stations above $40\,$m of distance to the reconstructed shower core with a signal above the expected one for a muon. This variable is then normalised to the total amount of signal encountered in stations far away from the shower core ($> 40\,$m);
\item \emph{distance} of the WCD station signal to the shape of the average gamma-ray lateral distribution function (LDF), also known as compactness.
\end{itemize}

 \begin{figure}
     \centering
     \includegraphics[align=c, width=.7\textwidth]{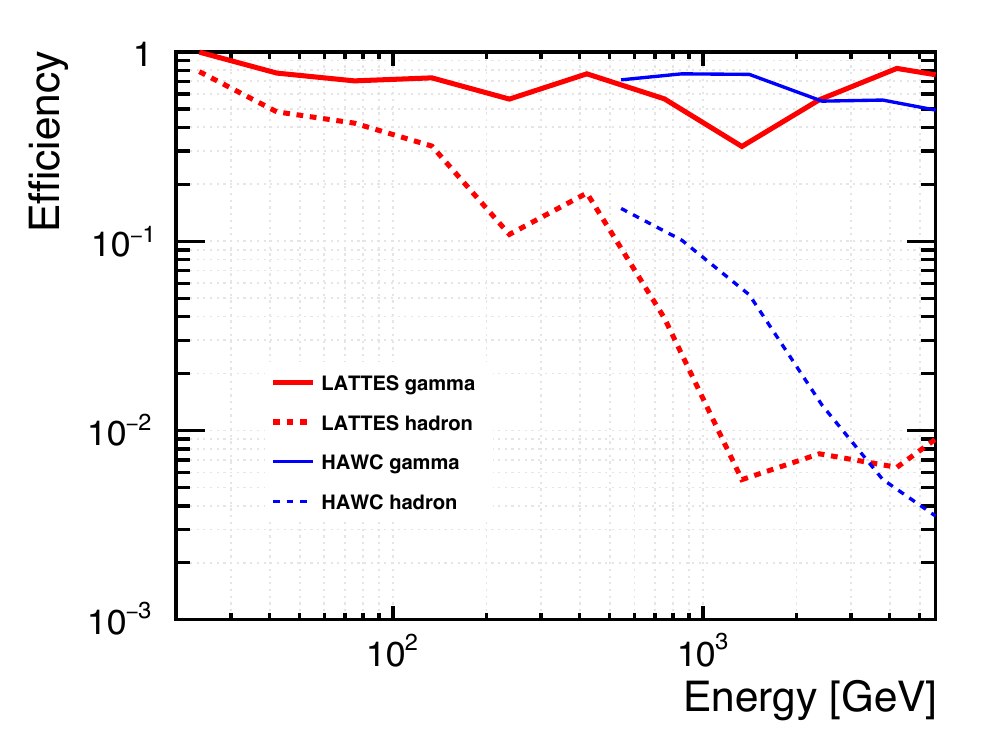}
     \caption{Gamma (solid lines) and hadron (dashed lines) selection efficiency as a function of the reconstructed energy. The HAWC curve is shown for comparison~\cite{HAWC2017}.}
  \label{fig:ghdisc}
\end{figure}

Both variables were combined using a Fisher linear discriminant and the gamma selection efficiency and hadron rejection results are presented in figure~\ref{fig:ghdisc}. As one can seen, this preliminary analysis gives results already as good as those reached by the HAWC collaboration.

\subsection{Sensitivity to steady sources}

\begin{figure}
     \centering
     \includegraphics[align=c, width=.7\textwidth]{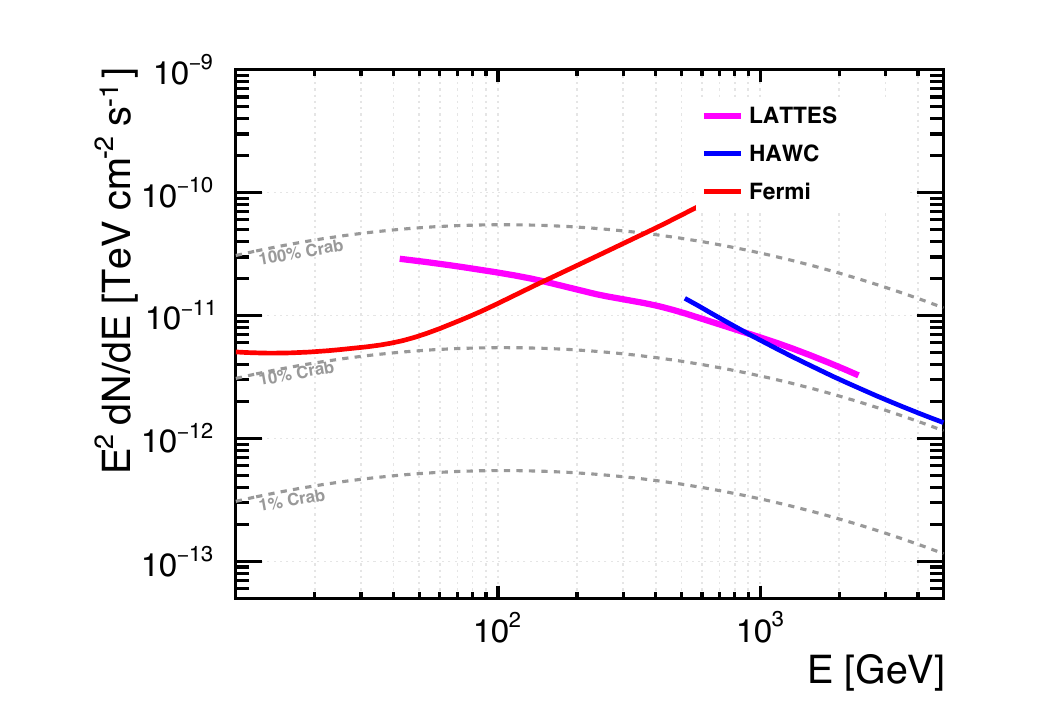}
     \caption{Differential sensitivity as a function of the reconstructed energy. We compute the flux of the source in a given energy range for which $N_{excess} / N_{bkg} = 5, N_{excess} > 10$, after 1 year of observation time (a 25\% duty cycle has been assumed). 5 bins per decade in estimated energy are used. For comparison, fractions of the Crab Nebula spectrum are plotted with the thin dashed gray lines.}
  \label{fig:sensitivity}
\end{figure}

Using the results described in the previous sections, one can now compute this detector sensitivity to steady sources. We compute the differential sensitivity as the flux of a source giving ${\rm N_{excess} /  \sqrt{N_{bkg}}} = 5$ after 1 year of effective observation time. It was assumed that the source is visible one fourth of the time. This is roughly the time that the galactic centre is visible in the Southern hemisphere. The results from this study are shown in figure~\ref{fig:sensitivity} and are compared with the 1 year sensitivities of FERMI and HAWC. One can clearly see that this detector would be able to cover the gap between the two of the most sensitive experiments in this energy range.

\section{Summary}

We have presented a novel hybrid detector which combines Resistive Plate Counters with a Water Cherenkov Detetor, able to extend the sensitivity of previous experiments down to the region of 100 GeV.

This modular compact and low cost detector, has been assessed with a realistic simulation, giving encouraging results. However, its capabilities are far from being fully explored. More sophisticated analysis are expected to be achieved in order to take fully advantage of LATTES hybrid concept. Studies to include a sparse array to extend the LATTES energy range up to 100 TeV are also undergoing.

Finally, with the advent of the Cherenkov Telescope Array, LATTES would be a fully complementary project that could provide not only triggers to transient phenomena but also contribute with long term observations of variable sources.

\bibliographystyle{jhep}
\bibliography{Biblio}

\providecommand{\href}[2]{#2}\begingroup\raggedright\begin{thebibliography}{1}

\bibitem{corsika}
D.~Heck, G.~Schatz, T.~Thouw, J.~Knapp and J.~N. Capdevielle, \emph{Corsika: A
  monte carlo code to simulate extensive air showers}, .

\bibitem{geant4}
S.~Agostinelli et~al., \emph{Geant4 -- a simulation toolkit}, {\emph{Nucl.
  Instrum. Meth.} {\bfseries A506} (2003) 250--303}.

\bibitem{HAWC2017}
A.~U. Abeysekara et~al., \emph{{Observation of the Crab Nebula with the HAWC
  Gamma-Ray Observatory}},
  \href{https://doi.org/10.3847/1538-4357/aa7555}{\emph{Astrophys. J.}
  {\bfseries 843} (2017) 39},
  [\href{https://arxiv.org/abs/1701.01778}{{\ttfamily 1701.01778}}].

\end{thebibliography}\endgroup

\end{document}